\documentclass[twocolumn,showpacs,prl,superscriptaddress]{revtex4}
\usepackage{graphicx}
\usepackage{color}

\begin{document}

\title{Generalized maximum entropy approach to quasi-stationary states in long range systems}

\author{Gabriele Martelloni}\email{gmartell@sissa.it}
\affiliation{SISSA, Via Bonomea 286, 34137 Trieste, Italy}
\author{Gianluca Martelloni}\email{gianluca.martelloni@unifi.it}
\affiliation{Department of Physics and Astronomy, CSDC and INFN University of Florence, via G. Sansone 1, 50019 Sesto Fiorentino, Firenze, Italy}
\author{Pierre de Buyl}\email{pierre.debuyl@chem.kuleuven.be}\thanks{P.d.B. is a Postdoctoral Fellow of the Research Foundation - Flanders (FWO).}
\affiliation{Instituut voor Theoretische Fysica, KU Leuven B-3001, Belgium}
\author{Duccio Fanelli}\email{duccio.fanelli@unifi.it}
\affiliation{Department of Physics and Astronomy, CSDC and INFN University of Florence, via G. Sansone 1, 50019 Sesto Fiorentino, Firenze, Italy}

\date{\today}

\begin{abstract}
  Systems with long-range interactions display a short-time relaxation towards Quasi Stationary States (QSS) whose lifetime increases with the system size. In the paradigmatic  Hamiltonian Mean-field Model (HMF) out-of-equilibrium phase transitions are predicted and numerically detected which separate homogeneous (zero magnetization) and inhomogeneous (nonzero magnetization) QSS. In the former regime, the velocity distribution presents (at least) two large, symmetric, bumps, which cannot be self-consistently explained by resorting to the conventional Lynden-Bell maximum entropy approach. We propose a generalized maximum entropy scheme which accounts for the pseudo-conservation of additional charges, the even momenta of the single particle distribution.  These latter are set to the asymptotic values, as estimated by direct integration of the underlying Vlasov equation, which formally holds in the thermodynamic limit. Methodologically, we operate in the framework of a generalized Gibbs ensemble, as sometimes defined in statistical quantum mechanics, which contains an infinite number of conserved charges. The agreement between theory and simulations is satisfying, both above and below the out of equilibrium transition threshold.
A precedently unaccessible feature of the QSS, the multiple bumps in the velocity profile, is resolved by our new approach.
\end{abstract}


\maketitle

Systems subject to long-range forces are known to display intriguing feature, which manifest both at equilibrium and out of equilibrium \cite{book, review}. As an example  starting from out--of--equilibrium initial conditions, long-range systems get trapped in long lasting Quasi-Stationary States (QSS), whose lifetime diverges with the system size. Importantly, when performing the mean-field limit {\it before} the infinite time limit, the system cannot relax towards the deputed Boltzmann--Gibbs equilibrium and stays permanently confined in the QSS. In this regime, the relevant order parameters take values distinct from those attained  at equilibrium  and the system exhibits a large gallery of peculiar anomalies, as e.g. non
Gaussian velocity distributions. QSS have been reported to occur for a wide plethora of physical systems for which
long-range couplings are at play. These include plasma wave instabilities \cite{Nicholson} relevant for fusion devices, self-gravitating systems \cite{Padmanabhan} invoked in the study of non baryonic large scale structures formation in the Universe, and Free Electron Laser \cite{Barre,Teles}, light sources of paramount importance for their intrinsic flexibility. To elucidate the mechanisms which drive the emergence of QSS proved a challenging task, that stimulated a vigorous, still open, debate.

QSS have been interpreted as Vlasov stable stationary states, which evolve because of the collisional, finite $N$, effects.  One can hence gain analytical insight into the QSS, by resorting to a Vlasov based maximum entropy scheme, as originally pioneered by Lynden-Bell in his seminal paper \cite{LyndenBell68}. The aforementioned method has been successfully applied both to paradigmatic toy models \cite{AntoniazziFanelli}  and real world systems  \cite{Barre}, allowing to unravel a rich zoology of out-of-equilibrium attributes that encompass phase transitions \cite{AntoniazziFanelli_tricrit} and phase re-entrances \cite{ChavanisFanelli}. Remarkably, not a single adjustable parameter is employed in the theory, which has therefore a fully predictive value. Despite the general agreement with the results of direct simulations, deviations are punctually detected when the ergodic hypothesis, which ultimately underlies the maximization procedure, fails. Dynamical effects need to be properly included into the theory so to eventually improve its inherent prognostic ability. Working along these lines, the so called core halo solution was proposed in \cite{Pakter}. The mechanisms of core-halo formation shares similarities to the process of evaporative cooling: macroscopic density waves propagates in the hosting medium. Particles matching the resonant conditions can gain energy at the detriment of collective modes and thus leave the inner core to populate a diffusing halo. On the other hand, the loss of energy drives a condensation into a low energy state of the particles constituting the bulk. Due to the Vlasov constraints, the core cannot freeze by eventually collapsing into the minimum of the potential. It in turn approaches the maximum allowed phase space density, as imposed by the assigned initial conditions. The method results in a semi-analytical strategy to accurately characterized the QSS regime in the regime where particles bunch inside the separatrix of a collective resonance (magnetized phase, as described below). Notwithstanding the undoubtedly success of the theories mentioned above, we are however far from having said the final word on the genesis of QSS. With reference to the celebrated Hamiltonian Mean Field (HMF) model \cite{AntoniRuffo}, the classically recognized  arena for challenging QSS studies, pronounced bumps have been for instance reported to spontaneously set in for the homogeneous QSS regime \cite{AntoniazziCalifano, BachelardPRL}. These vortices, or phase space clumps, imply that the relaxation is incomplete. No explanation of this macroscopic, collective phenomenon has been so far proposed. Starting from these premises, the aim of this Letter is to discuss a plausible modification of the original Lynden-Bell approach, so to reconcile theory and empirical observations. Although we will develop the analysis for the HMF model, the method is general and could be in principle mutuated to those settings, from plasma physics to Free Electron Laser, where the QSS have been shown to occur. 

Let us start by introducing the HMF model, which describes the coupled dynamics of $N$ classical rotators. The Hamiltonian reads:
\begin{equation}
\label{eq:ham}
H = \frac{1}{2} \sum_{j=1}^N p_j^2 + \frac{1}{2 N} \sum_{i,j=1}^N 
\left[1 -  \cos(\theta_j-\theta_i) \right]
\end{equation}
where $\theta_j$ stands for the orientation of the $j$-th rotor and
$p_j$ is its conjugate momentum. To monitor the evolution of the
system, one can rely on the magnetization, a global order
parameter defined as $M=|{\mathbf M}|=|\sum {\mathbf m_i}| /N$, where
${\mathbf m_i}=(\cos \theta_i,\sin \theta_i)$ represents the local
magnetization vector. As already recalled, starting from {\it out--of--equilibrium} initial
conditions, the system gets frozen in a QSS, whose time duration diverges with the number of simulated particles $N$. 
When the limit $N \rightarrow \infty$ is performed {\it before} the infinite time limit ($t \rightarrow \infty$)
the system is stuck in the QSS and cannot proceed towards its associated equilibrium configuration  \cite{Yamaguchi}.
In other words, QSSs can be interpreted as stationary stable equilibria 
of the system in its mean-field continuum representation. This observation immediately translates into a rigorous 
route to analytically inspect the peculiar out-of-equilibrium dynamics of systems subject to long-range couplings.
  
To this end, we preliminary recall that, in the $N \to \infty$ limit, the
$N$-particle dynamics of Hamiltonian~(\ref{eq:ham}) (similarly for Hamiltonian of the same class) yields the following Vlasov equation

\begin{equation}
\frac{\partial f}{\partial t} + p\frac{\partial f}{\partial \theta} -
\frac{d V}{d \theta} \frac{\partial f}{\partial p}=0\quad ,
\label{eq:VlasovHMF}
\end{equation}

where $f(\theta,p,t)$ denotes the microscopic one-particle
distribution function and
\begin{eqnarray}
V(\theta)[f] &=& 1 - M_x[f] \cos(\theta) - M_y[f] \sin(\theta) ~, \\
M_x[f] &=& \int_{-\pi}^{\pi} \int_{-\infty}^{\infty}  f(\theta,p,t) \, \cos{\theta}  {\mathrm d}\theta
{\mathrm d}p\quad , \\
M_y[f] &=& \int_{-\pi}^{\pi} \int_{\infty}^{\infty}  f(\theta,p,t) \, \sin{\theta}{\mathrm d}\theta
{\mathrm d}p\quad .
\label{eq:pot_magn}
\end{eqnarray}
The specific energy $h[f]=\int \int (p^2/{2}) f(\theta,p,t) {\mathrm d}\theta
{\mathrm d}p - ({M_x^2+M_y^2 - 1})/{2}$ and momentum
$P[f]=\int \int p f(\theta,p,t) {\mathrm d}\theta
{\mathrm d}p$ functionals are conserved quantities, together with mass normalization.

The main idea of the Lynden-Bell maximum entropy theory,
is to coarse-grain the microscopic one-particle distribution function
$f(\theta,p,t)$. It is then possible to
associate an entropy to $\bar{f}$, the coarse-grained version of the single-particle distribution.
The sought statistical equilibrium can be determined by maximizing such an entropy while imposing the
conservation of the Vlasov dynamical invariants \cite{LyndenBell68,Chavanis06, Michel94}.

We shall here assume that the initial distribution is of the {\it water-bag} type. We will in particular restrict the analysis to rectangular water-bag in the ($\theta,p$) plane, centered in the origin and of respective half widths $\Delta_{\theta}$ and  $\Delta_{p}$. The distribution $f$ takes therefore just two values:
$f_0=1/(4 \Delta_{\theta} \Delta_{p})$, inside the boundaries of the rectangle, and zero outside.
The water-bag geometry is entirely specified once the energy
$h[f]=e$, and the initial magnetization ${\mathbf
M_0}=(M_{x0}, M_{y0})$ are assigned.  

The Vlasov time evolution can reshape 
the boundaries of the water-bag, while preserving the area inside
it. The distribution stays therefore two-levels ($0,f_0$) as time
progresses. In this two-level
representation, the
mixing entropy per particle associated to $\bar{f}$ reads
\begin{equation}
\label{entropy_shape}
s(\bar{f})=-\int \!\!{\mathrm d}p{\mathrm d}\theta \,
\left[\frac{\bar{f}}{f_0} \ln \frac{\bar{f}}{f_0}
+\left(1-\frac{\bar{f}}{f_0}\right)\ln
\left(1-\frac{\bar{f}}{f_0}\right)\right].
\end{equation}
as it follows from a straightforward combinatorial 
analysis\cite{LyndenBell68,Chavanis06}.
The Lynden-Bell recipe amounts to solve the
following constrained variational
problem
\begin{eqnarray}
S(e,\sigma)\! =\! \max_{\bar{f}} \biggl(\!  s(\bar{f})
\biggr|
 h(\bar{f})=e;\;\!\! P(\bar{f})=\sigma;\; \!\!\!
\int \!\! {\mathrm d}\theta {\mathrm d}p \bar{f}=1\biggr)~.
\label{eq:problemevar}
\end{eqnarray}
To carry out the calculations one needs to introduce three Lagrange multipliers
$\beta/f_0$, $\lambda/f_0$ and $\mu/f_0$ for respectively energy, momentum and
normalization. This leads to the following analytical form of the
distribution
\begin{equation}
\label{eq:barf} \bar{f}(\theta,p)= f_0\frac{e^{-\beta (p^2/2
- M_y[\bar{f}]\sin\theta
- M_x[\bar{f}]\cos\theta)-\lambda p-\mu}}
{1+e^{-\beta (p^2/2  - M_y[\bar{f}]\sin\theta
 - M_x[\bar{f}]\cos\theta)-\lambda p-\mu}},
\end{equation}

which differs from the Boltzmann-Gibbs expression because of the
``Fermionic'' denominator that originates from the specific nature of the
entropy (\ref{entropy_shape}). By inserting expression (\ref{eq:barf}) into energy, momentum and
normalization constraints and by making use of the definition of
magnetization, one can obtain a set of implicit 
equations in the unknowns $\beta$, $\lambda$, $\mu$, $M_x$ and $M_y$. Such a system can be solved,
for any supplied values of the energy $e$ and the initial magnetization $M_0$,  to determine
the Lagrange multipliers, together with $M_x$ and $M_y$, for which the
Lynden-Bell entropy is extremal.

\begin{figure}[h]
\begin{center}
\includegraphics[width=8cm]{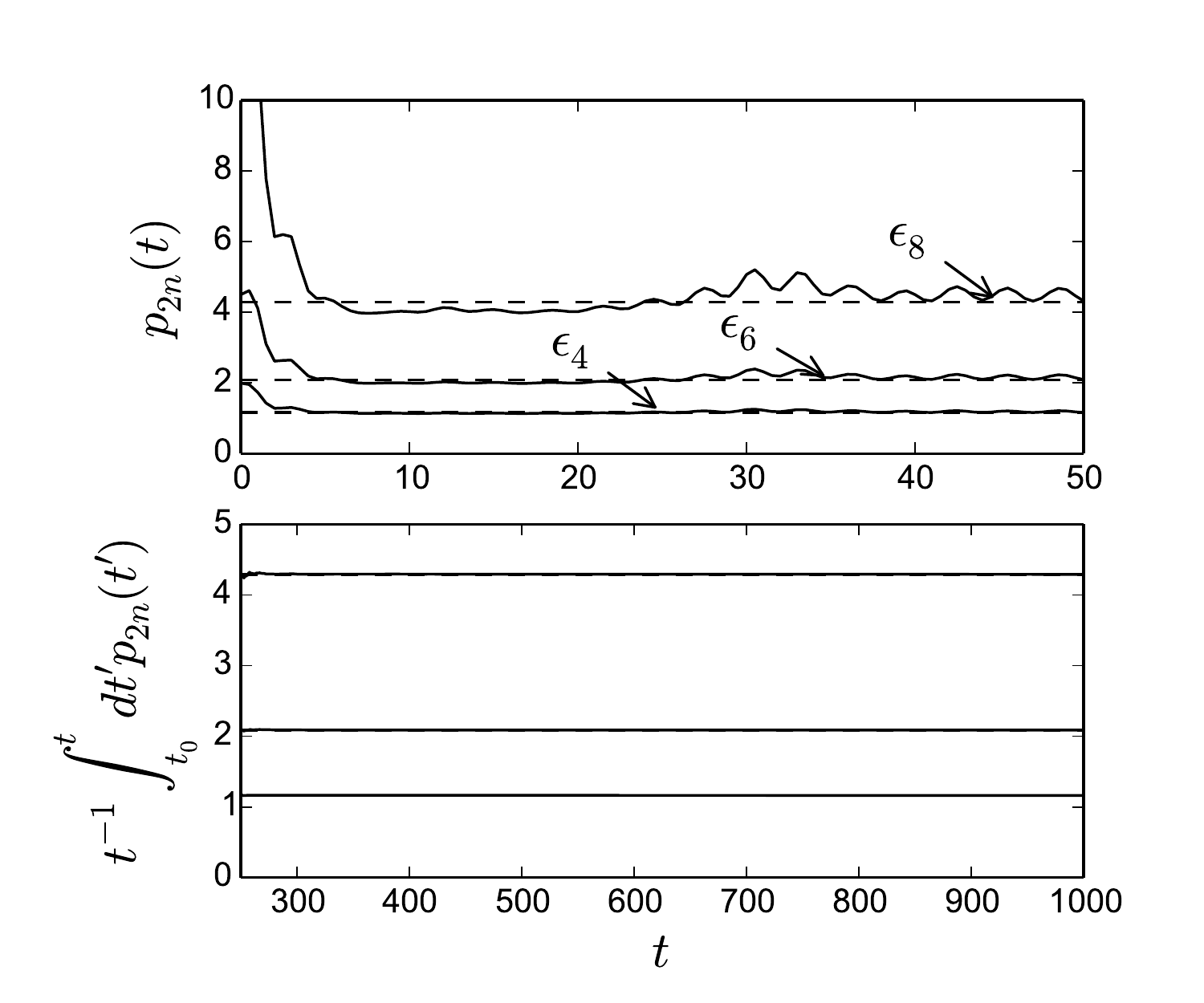}
\end{center}
 \caption{Upper panel: time evolution of the first few moments $p_{2n}$ with $n=1,2,3$. After a violent relaxation the moments converges to their asymptotic values $\epsilon_{2n}$. The results refer to a choice of the parameter which yields homogeneous QSS. Similar observation holds for the magnetized QSS phase. Lower panel: running averages in the time interval $(250,1000)$.}
\label{fig1}
\end{figure}

Interestingly, the calculation sketched above enabled one to detect an out-of-equilibrium phase transition  \cite{AntoniazziFanelli_tricrit}
in the reference plane $(M_0, e)$: when assigning the initial magnetization and decreasing the energy density $e$,
the system experiences a switch from a homogeneous (zero magnetization) to a non homogeneous (magnetized) QSS. Consequently, the plane of parameters $(M_0,e)$ can be partitioned into two adjacent zones, respectively associated to an ordered  and  disordered phase. The transition line, which separates the aforementioned regimes, is a collection of critical points, as determined by the Lynden-Bell theory. Both first and second order phase transition are identified, which merge together into a tricritical point. The details of the transition have been more recently revisited through the core-halo approach \cite{Pakter}, so resulting in a refined description of the scrutinized phenomenon.  Above threshold, when the system is predicted to relax towards a homogeneous QSS, two symmetric bumps appear in the velocity distribution, as recorded via $N$-body and Vlasov based simulations. This is the imprint of a collective effect
which yields the formation of two clusters in the $(\theta,p)$ plane. Neither the Lynden-Bell theory nor
the core-halo semi-analytic approach could provide a plausible account for the spontaneous rise of the bumps, an observation which remains to date unexplained. The aim of this Letter is to discuss an extension of the Lynden-Bell maximum entropy scheme, which accommodates for a set of additional Vlasov constraints that prove necessary to benchmark theory and simulations. We will here exclusively report numerical simulations of the Vlasov equation, the continuum counterpart of the original $N$-body Hamiltonian (\ref{eq:ham}).
All simulations were performed with the vmf90 program \cite{vmf90}.
Provided $N$ is sufficiently large, discrete $N$-body simulations and their corresponding Vlasov homologue are virtually identical \cite{AntoniazziCalifano}. Incidentally, we will also show that the velocity profiles recorded above threshold return a rather intricate gallery of possible structures, ranging from two to multiple bumps, an observation which contributes to significantly enrich the phenomenology so far reported in the literature \footnote{The hypothetical possibility of obtaining exotic features of the velocity distribution function from the inclusion of additional, exactly conserved invariants, has been also considered in an abstract setting in \cite{DeBuyl}}.
\\
Consider the $2n$-th moment $\langle p^{2n} \rangle$ of the distribution function defined as
$p_{2n} \equiv \langle p^{2n} \rangle = \int_{-\pi}^{\pi} \int_{-\infty}^{\infty} p^{2n} f(\theta, p) d \theta d p$. For obvious symmetry reasons, given the chosen family of initial conditions, only even moments need to be considered, the odd ones being identical to zero.  Starting from out of equilibrium initial condition of the water-bag type, with specified energy and initial magnetization, one can follow the time course of $\langle p^{2n} \rangle$ for arbitrary choices of the even integer $n$. The typical result of a large campaign of simulations is depicted in figure \ref{fig1}): after a violent relaxation, as it is customarily termed the sudden initial evolution, the moments approach stationary plateaux characterized by well defined average values. In formulae, $\lim_{t \rightarrow \infty } p_{2n} = \epsilon_{2n}$, where $\epsilon_{2n}$ are scalars determined by the simulations. The Vlasov dynamics is sampling its equilibrium solution: the modest oscillations displayed by  $p_{2n}$ around their mean values  $\epsilon_{2n}$, as shown in figure \ref{fig1}, reflect the local wandering of the trajectory inside the target basin of attraction. In the discrete realm, finite $N$ effects are indeed responsible for the subsequent slow evolution towards the asymptotic Boltzmann equilibrium. Working in the continuum setting implies silencing endogenous finite $N$ corrections and consequently preventing the system from leaving the collision-less Vlasov equilibria. These are the QSS that we aim at characterizing and which should be self-consistently coerced to match the average values for the even moments  $p_{2n}$, as measured from Vlasov simulations. Building on this observation, we elaborate on a possible modification of the Lynden-Bell maximum entropy solution, which is constrained to reproduce the even moments of the distribution, treated as additional pseudo-conserved quantities.  The moments themselves are not Vlasov invariant, but they contribute as dynamical constraints to effectively delimit the portion of available phase-space. In the following, for purely demonstrative purposes, we will make use of the average values for $\langle p^n \rangle$, as resulting from the simulations. In general it is however possible to construct dedicated moments closure scheme which enables to estimate the sought quantities, so yielding to a fully predictive approach to the QSS characterization \cite{chandre}.

Mathematically, we shall study the generalized variational problem 

\begin{eqnarray}
S(e,\sigma, \epsilon_{2n})\! &=&\! \max_{\bar{f}} \biggl(\!  s(\bar{f})
\biggr|
 h(\bar{f})=e;\;\!\! P(\bar{f})=\sigma;\; \!\!\!\nonumber\\
&&\int \!\! {\mathrm d}\theta {\mathrm d}p \bar{f}=1;\;p_{2n}=\epsilon_{2n} \biggr)~.
\label{eq:problemevar}
\end{eqnarray}

By introducing the Lagrange multipliers $\nu_{2n}/f_0$ to account for the additional imposed constraints and performing a straightforward calculation one end up with the following expression for the extremal distribution function $f_{QSS}(\theta,p) \equiv \bar{f}(\theta,p)$:

\begin{equation}
\label{eq:barf_mod} {f}_{QSS}= f_0\frac{e^{-\beta (p^2/2
- M_y[\bar{f}]\sin\theta
- M_x[\bar{f}]\cos\theta)-\sum_{n} \nu_{2n} p^{2n} - \mu}}
{1+e^{-\beta (p^2/2  - M_y[\bar{f}]\sin\theta
 - M_x[\bar{f}]\cos\theta)-\sum_{n} \nu_{2n} p^{2n}-\mu}}.
\end{equation}

\begin{figure}[h]
\begin{center}
\includegraphics[width=7cm]{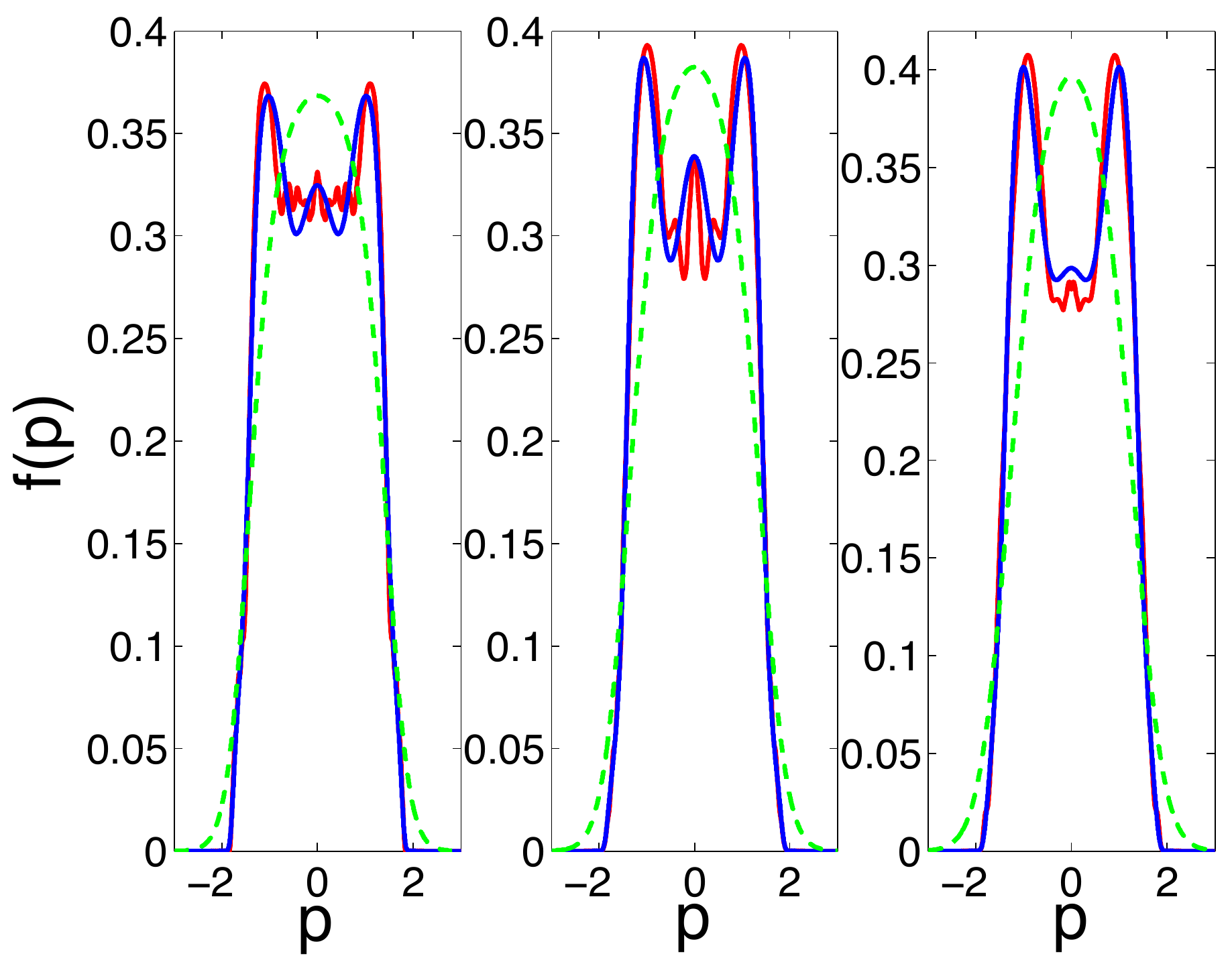}
\end{center}
\caption{The velocity distribution profiles are displayed for $e=0.9$ and different choices of the initial magnetization (from left to right, $M_0=0.3,0.5,0.7$). In all cases the system is predicted to converge to a homogeneous QSS. The result of the Vlasov simulations are depicted with a (red online) solid line. The standard Lynden-Bell theory yields the (green online) dashed curve. The modified maximum entropy scheme with the inclusion of the additional pseudo conserved quantities yields the profiles represented with (blue online) solid lines.}
\label{fig2}
\end{figure}

Notice that we have set $\lambda=0$, as it immediately follows for symmetry reason. The $n+2$ Lagrange multipliers, together with the unknown magnetization amount, can be determined by imposing the selected constraints. These are the conserved quantities, energy and normalization, and the stationary values for the even moments of the  distribution:

\begin{eqnarray}
\label{eq:cond0}
&&\int_{-\pi}^{\pi} \int_{-\infty}^{\infty}   {f}_{QSS} d \theta dp = 1 \nonumber \\
&&\int_{-\pi}^{\pi} \int_{-\infty}^{\infty}   p^2 {f}_{QSS}  d \theta dp =  e+\frac{M^2 - 1}{2} \nonumber\\ 
&&\int_{-\pi}^{\pi} \int_{-\infty}^{\infty}  p^{2n} {f}_{QSS}  d \theta dp = \epsilon_{2n} \qquad \forall n \ge 2 \nonumber \\
&&\int_{-\pi}^{\pi} \int_{-\infty}^{\infty} \cos(\theta)   {f}_{QSS}  d \theta dp = M_x \nonumber \\ 
&&\int_{-\pi}^{\pi} \int_{-\infty}^{\infty} \sin(\theta)   {f}_{QSS} d \theta dp = M_y 
\end{eqnarray}
where we recall that ${f}_{QSS}$ depends self-consistently on ($\beta, \mu, \nu_{2n}, M_x,M_y$) as dictated by relation (\ref{eq:barf_mod}). In agreement with the original Lynden-Bell scheme, the generalized conditions (\ref{eq:cond0}) predicts
a parameter space $(M_0,e)$ partitioned into two region, corresponding to magnetized and homogeneous QSS. Above threshold, when the QSS magnetization is about zero, equation (\ref{eq:barf_mod}) reduces to the following compact expression, only function of $p$ (and so called velocity distribution):  

\begin{equation}
\label{eq:barf_mod} {f}_{QSS}(p)= f_0\frac{e^{-\beta p^2/2
-\sum_{n} \nu_{2n} p^{2n} - \mu}}
{1+e^{-\beta p^2/2  -\sum_{n} \nu_{2n} p^{2n}-\mu}},
\end{equation}

which can be interpreted as an asymptotic free theory. The force field acting during the violent relaxation phase is implicated in the short time modulation of $\langle p^{2n} \rangle$, and indirectly enter the picture through the 
quantities $\epsilon_{2n}$. In this respect, the proposed approach to the QSS resembles 
a quantum quench from interacting to free theory, as for instance discussed in \cite{SotiriadisCalabrese}, or, more recently, in \cite{SotMarte} where the inspected system is shown to equilibrate to a Generalised Gibbs Ensemble \cite{rdyo07}. In figure \ref{fig2} the velocity distribution $f(p)= \int f(\theta,p) d \theta$ as recorded from Vlasov based simulation is plotted (red solid line) in the region of a homogeneous QSS, for different values of the initial magnetization. The obtained profiles display the two characteristic bumps to which we alluded above. Other collective structures are however present and become more evident when $M_0$ is made larger, for fixed energy amount (right most panel).

\begin{figure}[h]
\begin{center}
\includegraphics[width=7cm]{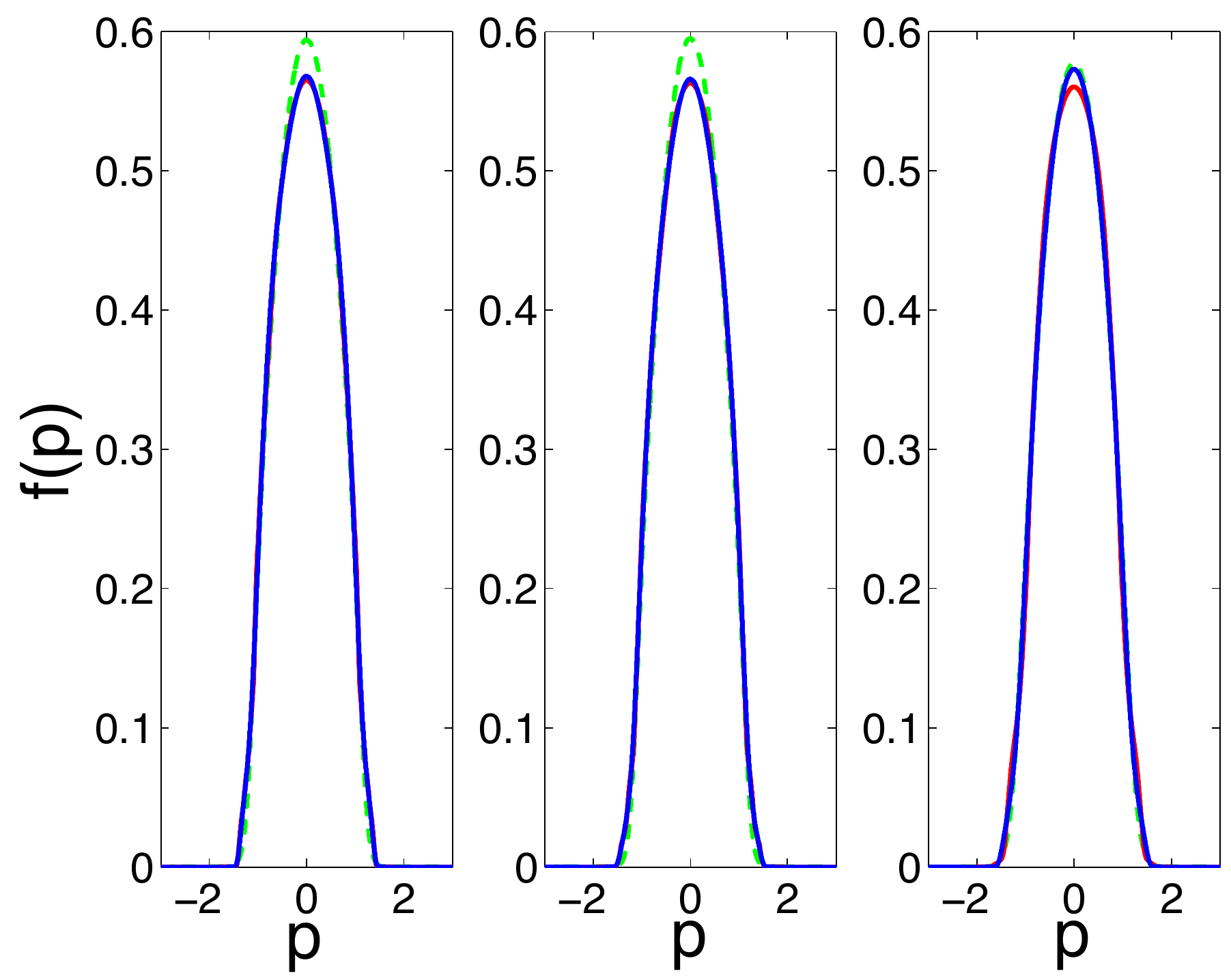}
\end{center}
\caption{The velocity distribution profiles are plotted for $e=0.6$ and different choices of the initial magnetization (from left to right, $M_0=0.3,0.5,0.7$). The system converges to a magnetized QSS. The result of the Vlasov simulations are drawn with a (red online) solid line. The standard Lynden-Bell theory yields to the (green online) dashed curve. Predictions based on modified maximum entropy scheme with the inclusion of the additional pseudo conserved quantities are depicted with  (blue online) solid lines. These latter are almost identical to the simulated profiles on the first two panels, from left to right.}
\label{fig3}
\end{figure}

The velocity distribution predicted by the standard Lynden-Bell theory (dashed green) fails to capture the fine details of the simulated profile. At variance, the generalized theory here described, see eq. (\ref{eq:barf_mod}),  which sets the even average momenta to the asymptotic values determined by Vlasov dynamics, proves definitely more adequate in explaining the observations (solid blue). The macroscopic details of the velocity profile are adequately captured upon truncation at the tenth order ($n=5$) in the hierarchy of pseudo conserved momenta. Notice that the improvements are evident not only in the bulk of the distribution, where the resonant particles are trapped, but also in the tails, the theoretical curves adhering better to the corresponding numerically computed lines. Smaller coherent structures appear in the numerics which are not captured by the theory, at this level of resolution. By accounting for higher moments ($n>5$) or
additional pseudo conserved charges might contribute to ameliorate the agreement even further. 
The inclusion of additional dynamical constraints is also beneficial in the region of magnetized QSS, as it can be appreciated by visual inspection of figure  
(\ref{fig3}).

In conclusion we have here discussed a generalization of the maximum entropy principle due to Lynden-Bell and focus on explaining the emergence of Quasi Stationary States in systems subject to long-range forces. By imposing additional dynamical constraints, stemming from the Vlasov equation that rules the evolution of the system in the continuum limit, allowed us to considerably enhance the predictive ability of the theory. In particular, by constraining the asymptotic values of the even moments of the velocity distribution, we theoretically demonstrated that the existence of two (or multiple) islands in the single particle phase-space is compatible with a maximum entropy principle.
Our conclusion is twofold. On the one side, we confirmed that a statistical mechanics treatment based on the governing Vlasov equation can be successfully invoked to address QSS peculiarities. Discrepancies between empirical observation and the standard Lynden-Bell theory can be significantly reduced, by taking explicit note of key dynamical restrictions. On the other side, we expect that our results could open up novel avenues to investigate those systems, of both theoretical or experimental relevance, where long range forces are active and QSSs have been observed (e.g. plasmas, Coulomb systems).

D.F. acknowledges financial support from the program PRIN 2012 financed by the Italian MIUR.
Gabriele Martelloni acknowledges the ERC for financial support
under Starting Grant 279391 EDEQS. The authors thanks P. Di Cintio and S. Ruffo for stimulating discussions.

\end{document}